\documentclass[nofootinbib,amsmath,amssymb,aps,prl,reprint,noeprint,floatfix,superscriptaddress,longbibliography]{revtex4-2}

\usepackage{float}
\usepackage{graphicx}
\usepackage{dcolumn}
\usepackage{soul} 

\usepackage{bm}
\usepackage[version=4]{mhchem}
\usepackage[breaklinks,colorlinks,linkcolor={blue}, %
            citecolor={blue},urlcolor={blue}]{hyperref}
\usepackage{siunitx}
\usepackage{physics}  
\usepackage{amssymb}
\usepackage{amsfonts}                  
\usepackage{graphicx}         
\usepackage[T1]{fontenc}     
\usepackage{ae}                     
\usepackage{lastpage}         
\usepackage{textcomp}
\usepackage{tabularx}
\usepackage{xcolor}
\usepackage{color}
\usepackage{nicefrac}
\usepackage{upgreek}
\usepackage{float}
\usepackage{mathtools}
\usepackage{gensymb}
\usepackage{pdfcomment}

\usepackage[normalem]{ulem}  

\newcommand{\kzpo}{K$_2$Zr(PO$_4$)$_2$ }

\begin{document}

\title{Hidden chiral signatures in ferroaxial K$_2$Zr(PO$_4$)$_2$}

\author{Nora Taufertsh{\"o}fer}
\affiliation{Materials Theory, ETH Z{\"u}rich, 8093 Zurich, Switzerland}
\author{Awadhesh Narayan}
\affiliation{Solid State and Structural Chemistry Unit, Indian Institute of Science, Bangalore 560012, India}
\affiliation{Materials Theory, ETH Z{\"u}rich, 8093 Zurich, Switzerland}
\author{Nicola A. Spaldin}
\affiliation{Materials Theory, ETH Z{\"u}rich, 8093 Zurich, Switzerland}

\date{\today}

\begin{abstract}

We use first-principles calculations and multipole analyses to demonstrate the relationship between ferroaxiality and chirality in the prototypical ferroaxial material K$_2$Zr(PO$_4$)$_2$. 
Using the atomic-site electric toroidal monopole as a measure of electronic chirality in real space, we show that, while the paraxial phase of \kzpo is non-chiral, the ferroaxial phase is antiferro-chiral. 
By applying an electric field, we induce a ferri-chiral state with net electronic chirality which is opposite for opposite underlying ferroaxial domains and can be tuned by the direction and strength of the electric field. 
Associated with the real-space induced chirality, we find a distinct response in momentum space, with induced non-zero components in the Berry curvature dipole tensor that switch sign between opposite ferri-chiral domains. 
Our findings therefore reveal hidden chirality in ferroaxial materials in both real and momentum space.
\end{abstract}

\maketitle

\section{Introduction}

Ferroaxial materials are a class of ferroic materials that preserve both time-reversal symmetry (TRS) and inversion symmetry (IS). 
As a result, the ferroaxial domains are unaffected by depolarizing magnetic or electric fields, making them appealing for technological use; conversely, they are challenging to manipulate systematically. Recently, ferroaxials have attracted increased interest following the prediction~\cite{He_Khalsa_PRR_optical_control_2024} and demonstration~\cite{Zeng_2025_switching} of switching of the domain structure by an engineered conjugate field using a circularly polarized terahertz light pulse that drives an infrared-active phonon.\\
In prototypical ferroaxial materials studied in the literature~\cite{Jin_RbFeMoO42:2020, Hayashida_domains_ferroaxials:2021, Nagai:2023}, the ferroaxial phase transition is driven by a rotational distortion. 
The glaserite compound K$_2$Zr(PO$_4$)$_2$ is a typical example of a pure ferroaxial material~\cite{Hlinka_PRL:2016} where the $\sim700$\,K transition is characterized by rotation of corner-sharing PO$_4$ tetrahedra and ZrO$_6$ octahedra~\cite{Yamagishi_KZPO:2023}. 
A suitable order parameter for the ferroaxial phase transition is the electric toroidal (ET) dipole moment~\cite{Dubovik:1986, Hayami_ferroaxial_moment:2022, Bhowal_2024_ETD} $\mathbf{G}$, an axial vector  that is invariant under spatial and time inversion.
The ET monopole $G_0 = \mathbf{G} \cdot \mathbf{q}$, a pseudoscalar quantity, has been proposed as the order parameter for chirality~\cite{Oiwa_PRL:2022, Inda_ETM:2024}, where $\mathbf{q}$ is a polar vector that introduces the necessary IS breaking. From this symmetry perspective, ferroaxiality has been proposed to be a precursor of chirality~\cite{Bhowal_2024_ETD,spaldin2026towards} and ferroaxial materials serve as a playground for studying hidden chiral features and correlations between polar and chiral order~\cite{Hayashida_2020_elecGyr, Hayami_chiral_axial_coupling_2025, Fava_PRL:2025}.\\
Complementary to the real space order parameters, the effects of chirality can also manifest in momentum space. In this regard, the Berry curvature (BC) -- which resembles a magnetic field in reciprocal space -- and its multipoles are especially characteristic~\cite{Xiao_Berry:2010, vanderbilt2018berry, sodemann2015quantum}. For example, the band dispersion of topological semimetals in chiral crystals is linked to the handedness of the structure~\cite{schroter2019chiral}, which results in an orbital-momentum locking~\cite{yang2023monopole}. However, the manifestations of BC and its multipoles in ferroaxial materials remain unexplored. \\
In this work, we use first-principles calculations and multipole analyses to investigate the relationship between ferroaxiality and chirality in both real and reciprocal space for the prototypical purely ferroaxial material K$_2$Zr(PO$_4$)$_2$ in the following termed KZPO. First, we show that the ferroaxial phase of KZPO can be considered to be antiferro-chiral, based on atomic-site ET monopole moments. Next, we induce structural chirality using an electric field. This generates a tunable net $G_0$ moment that switches sign between opposite ferroaxial domains for the same field orientation.
Lastly, we explore how electric fields can be used to manipulate the Berry curvature dipole (BCD) components 
by inducing opposite chirality in opposite ferroaxial domains and comment on the resulting non-linear Hall effect.

\section{Computational methods}

To perform the electronic structure calculations, we use density functional theory (DFT) within the local density approximation, as implemented in the Vienna ab initio simulation package (VASP)~\cite{Kresse:1996,Kresse:1999}. We use a $10\times10\times6$ $k-$grid for the high- and low-temperature unit cells of KZPO shown in Fig.~\ref{fig_struct_bands}a), an energy cutoff of 550 eV for the plane wave basis, and include spin-orbit coupling (SOC). The lattice parameters and internal coordinates are relaxed such that the forces acting on each atom are less than 1 meV/Å.\\ 
To study the onset of chirality, we introduce a small IS-breaking distortion that corresponds to the linear ionic response to an external electric field. 
Following ~\cite{Iniguez:2008},
we compute Born effective charges and the force-constant matrix that yields the infrared-active modes using density functional perturbation theory. We determine the atomic displacements due to a small electric field for both ferroaxial domains and the high-temperature structure.  
The maximum atomic displacement is 0.05 \AA{} for potassium at the largest applied field of $0.1~\text{V/\AA}$ along $z$, which keeps the structures close to the undistorted, inversion-symmetric reference and within the linear regime.\\ 
For the multipole analysis, we follow the method described in Refs.~\cite{Crichhio_PRL_multipole:2009, Bultmark_multipole:2009, VMueller_multipoles:2026} and decompose the density matrix $\rho_{lm, l'm'}$ obtained from DFT into irreducible spherical tensor components on atomic sites $i$ that are subsequently transformed into a real basis $w(l,l')^{\nu,kpr}_{t,i}$. Here, the indices $l$ and $l'$ refer to the orbital angular momentum channel, $\nu=$ 0 or 1 indicates a time-reversal even or odd multipole, $k$ is the orbital index, $p$ is the spin index, $r$ is the rank with \mbox{$r = [|k-p|,|k-p|+1 ,..., k+p]$}, and \mbox{$t=[-r, -r+1,...,r]$} is the component of the tensor. For example, for a rank $r=1$ tensor (i.e. a vector quantity), $t$ refers to the cartesian coordinates with $t=-1,0,1$ the $y, z, x$ component.\\ 
For the ET dipole moment, which acts as the order parameter for ferroaxiality, we extract the rank-1 tensor moment $\mathbf{w}(l,l')_{i}^{0,111} \propto \langle \mathbf{l} \times \mathbf{s} \rangle_i$~\cite{Hoshino_PRL:2023}, with $l+l'$ even from our multipole decomposition. 
As chirality measure, we consider the atomic-site ET monopole moment
\begin{align}
    G_{0,i} = \mathbf{G}_i \cdot \mathbf{p}_i 
    = \mathbf{w}(l,l')_i^{0,111} \cdot \mathbf{w}(l,l')_i^{0,101},
    \label{G0_def_G_p}
\end{align}
where $\mathbf{p}_i$ with $l+l'$ odd is the atomic electric dipole (ED) moment.\\
To calculate the properties related to the BC, we build a tight-binding model based on our DFT calculations using the method of maximally localized Wannier functions as implemented in \textsc{wannier90}~\cite{wannier90_Pizzi:2020}. 
The BC is given by
\begin{align}
    \Omega^n_a(\mathbf{k}) = -2 \hbar^2 \text{Im} \sum_{m\neq n} 
    \frac{\bra{n}\hat{v_b}\ket{m} \bra{m} \hat{v_c}\ket{n}}
    {(\epsilon_n-\epsilon_m)^2}
\end{align}
with $\hat{v_a}=\partial_{k_a} \hat{H}/\hbar$ the velocity operator, $\ket{m}$ and $\epsilon_m$ denoting the eigenstates and energy eigenvalues of the Hamiltonian $\hat{H}$ and $\hbar$ the reduced Planck's constant. The indices $a,b,c \in \{x,y,z \}$ refer to the direction. The BCD is then calculated from
\begin{align}
    D_{ab} = - \int_{\text{BZ}} d \mathbf{k} \sum_n 
    \biggl ( \partial_{k_a} f_0(\epsilon_{n \mathbf{k}} )\biggr )
    \Omega^n_{b}(\mathbf{k}) 
\label{BCD_Fsurf}
\end{align}
where $f_0$ is the equilibrium Fermi-Dirac distribution and BZ refers to the first Brillouin zone.
We use the implementation in \textsc{WannierBerri}~\cite{Tsirkin_WBerri:2021} to compute these quantities. Further details on the methods used are given in the Supplementary Material (SM).

\section{Structural and electronic properties of ferroaxial KZPO}

\begin{figure}[t]
\begin{center}
\includegraphics[trim = 0mm 72mm 0mm 0mm, clip, width=1\linewidth]{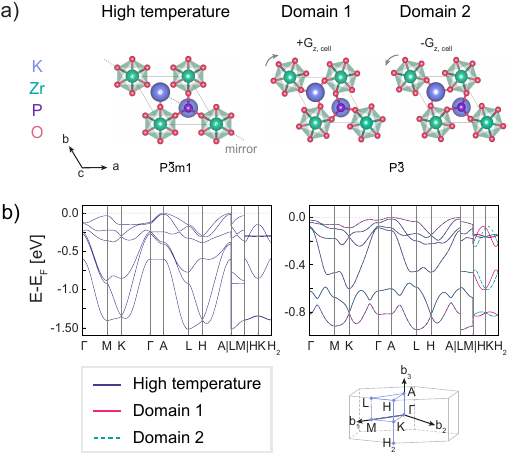}
\caption[]{Structural and electronic properties defining the ferroaxial phase transition in KZPO. a) High temperature and ferroaxial domain structures characterized by opposite net ET dipole moments $\pm G_{z, \text{cell}}$, visualized with \texttt{vesta}~\cite{VESTA_Momma:2008}. b) DFT band structures of the high temperature (left) and the ferroaxial (right) phase including SOC. The broken mirror symmetry results in different band dispersions for the two domain states along the edge of the BZ which is parallel to the rotational axis $z$ and hosts 3-fold rotational symmetry. The BZ with the high-symmetry points marked is shown at the bottom.}
\label{fig_struct_bands}
\end{center}
\end{figure}

In KZPO, the ferroaxial rotational distortion breaks the mirror symmetry of the high temperature space group P$\bar{3}$m1 to P$\bar{3}$, resulting in the ferroaxial phase. The rotational sense of the distortion distinguishes the two possible domain states visualized in Fig.~\ref{fig_struct_bands}a). 
As demonstrated in~\cite{Bhowal_2024_ETD}, the ET dipole moment defined as $\mathbf{G} \propto \langle \mathbf{l} \times \mathbf{s} \rangle_i$, which corresponds to the time-reversal and inversion even moment $\mathbf{w}^{0,111}$ in our multipole decomposition, is a suitable order parameter to describe the ferroaxial phase transition. 
Here, we compute these atomic-site moments for $l=l'=1$ and find that the dominant contributions come from the oxygen atoms at Wyckoff position 6g and the two phosphorus atoms. We define a net moment $\pm G_{z, \text{cell}}$ by taking the sum of $\mathbf{w}_i^{0,111}$ over all atoms $i$ in the unit cell and find that it has opposite sign in the two ferroaxial domains visualized in Fig.~\ref{fig_struct_bands}a).
Consistent with a recent report that the charge hexadecapole contributes to the ferroaxial phase transition~\cite{Xie_KZPO}, we find that the charge hexadecapole component $w_{3}^{0,404}$ with $l=l'=2$ on zirconium also changes sign between domains like the ET dipole moment.\\
Fig.~\ref{fig_struct_bands}b) shows the DFT band structures of the valence states close to the band gap of $\sim 4.3$ eV for the high- and  $\sim 4.9$ eV for the low-temperature phases including SOC. The energy zero is set to the top of the valence bands. 
Due to TRS and IS, Kramers' degeneracy is present at every $k$-point in the BZ. Without SOC, there are 4-fold degenerate bands with shallow dispersion along the high-symmetry paths which are parallel to the $z$ direction and host 3-fold rotational symmetry ($\Gamma-A$ and $H-K$). 
SOC splits these bands into $2+2$ Kramers' pairs. Interestingly, the band dispersions differ between the two ferroaxial domains along the $H-K$ path, reflecting the broken mirror symmetry compared to the high-temperature phase.

\section{Antiferro-chirality in real space}
\begin{figure}[t!]
\begin{center}
\includegraphics[trim = 0mm 110mm 0mm 0mm, clip, width=1\linewidth]{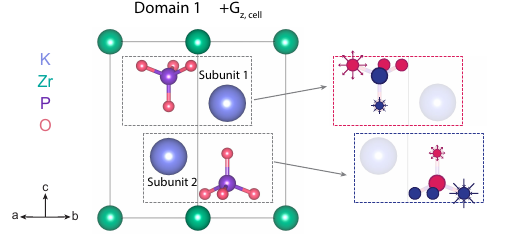}
\caption[]{Unit cell of KZPO for domain 1 with IS-breaking subunits indicated (left). These show an antiferroic pattern of local ET monopole moments, $G_0$. Positive (negative) $G_0$ moments are visualized by red (blue) and outward (inward) pointing 'hedgehogs' on the atomic sites. The relative area indicates the size of the $G_0$ moments.
}
\label{fig_subunits}
\end{center}
\end{figure}

While the crystal structure of KZPO is symmetric under spatial inversion, the unit cell consists of two individually inversion-breaking subunits~\cite{Bhowal_2024_ETD}, as depicted in~Fig.~\ref{fig_subunits} for domain 1.
From a symmetry perspective, these subunits are structurally chiral and therefore may host chiral signatures in the electron density. In the following, we show that KZPO is antiferro-chiral based on its local multipole moments.\\
As a measure for electronic chirality, we compute the ET monopole on each atomic site $i$ as given in eq. (\ref{G0_def_G_p}), with extracted $\mathbf{G}_i=\mathbf{w}(1,1)_{i}^{0,111}$ and $\mathbf{p}_i = \mathbf{w}(1,0)_{i}^{0,101}$ from our multipole decomposition.
We obtain finite local $G_{0}$ moments on the phosphorus and oxygen atoms. 
We find that the atoms in opposite subunits that are related by inversion carry opposite ET monopole moments (Fig.~\ref{fig_subunits}, right).
Therefore, while each subunit is chiral, the ET monopole moments cancel overall. For domain 2, all local $G_{0}$ moments are exactly opposite to domain 1. Note also that for the high-temperature paraxial phase, the ET monopoles vanish on all atomic sites.\\
%
%
\section{Induced ferri-chirality in real space}
\begin{figure}[b!]
\begin{center}
\includegraphics[trim = 0mm 67mm 0mm 0mm, clip, width=1\linewidth]{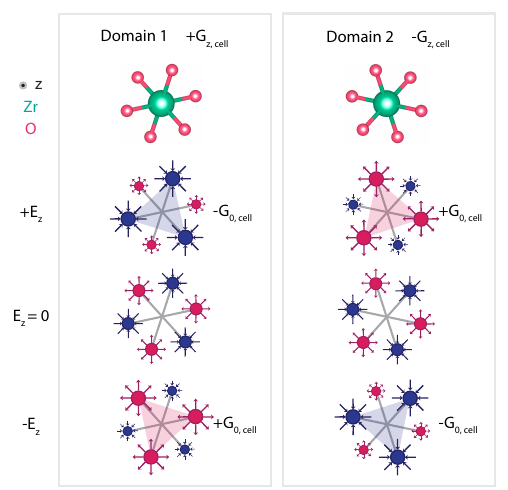}
\caption[]{Pattern of local $G_0$ moments in the ZrO$_6$ octahedra for both ferroaxial domains. The relative size and sign of the finite $G_0$ moments on the oxygen sites are schematically illustrated by the area and color of the 'hedgehogs'. While for zero electric field the ET monopoles render an antiferro-chiral state and cancel overall, a finite electric field (second and fourth row) imposes a disproportion between positive and negative moments leading to an induced ferri-chiral pattern and hence non-zero electronic chirality for the unit cell.}
\label{fig_local_G0}
\end{center}
\end{figure}
\begin{figure}[t!]
\begin{center}
\includegraphics[trim = 0mm 83mm 0mm 0mm, clip, width=1\linewidth]{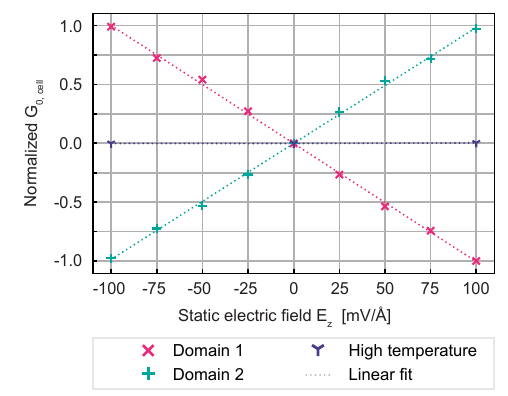}
\caption[]{Summed atomic ET monopoles over a unit cell as a function of electric field along $z$. For the same electric field, the net $G_{0,\text{cell}}$ is opposite for opposite ferroaxial domains, similar to opposite chiral enantiomers. Within one domain, $G_{0,\text{cell}}$ can be switched by reversing the electric field.}
\label{fig_net_G0}
\end{center}
\end{figure}

Now, we displace the ions according to the action of an IS-breaking applied electric field, which, from symmetry analysis, will introduce structural chirality. 
We apply fields up to $0.1$ V/\AA\, along the ferroaxial axis, here $z$. This direction is special as it breaks IS while keeping the 3-fold rotational symmetry, inducing a structural change from the space group P$\bar{3}$ to P3 and retaining the mirror relation between the two ferroaxial domains. Applying the field in other directions reduces the symmetry to P1 and eliminates the mirror relation between domains.\\
First, we examine again the local $G_{0,i}$ moments on each atom for both ferroaxial domains, this time for a finite applied electric field of $E_z=\pm 0.1$ V/\AA.
Depending on the direction of the electric field, the local ED moments change such that an imbalance is induced between the positive and negative local $G_{0,i}$ moments. This is illustrated in Fig.~\ref{fig_local_G0} for the ZrO$_6$ octahedra in both ferroaxial domains. From a microscopic viewpoint, the finite electric field biases the system into a state with positive or negative net electronic chirality by modifying the formerly antiferro-chiral state into a ferri-chiral state.\\  
Next, we demonstrate that the electric field strength can be used to tune the size of the net electronic chirality, by evaluating the sum over all atoms in the unit cell $G_{0,\text{cell}}= \sum_i G_{0,i}$ for a range of electric fields in the $z$ direction. As shown in Fig.~\ref{fig_net_G0}, the net moment, normalized by the largest absolute value of domain 1, is linearly susceptible to the electric field that imposes structural chirality. 
For the same applied field, $G_{0,\text{cell}}$ switches sign between opposite ferroaxial domains, and within one domain, the opposite direction of the field leads to opposite chirality. For zero field, the chirality measure vanishes as discussed in the previous section. The high-temperature structure does not exhibit electronic chirality even for finite electric fields since its ET dipole moment is zero. Therefore, IS breaking does not induce an ET monopole moment. Finally, we note that the electric-field-induced chirality described here is likely the origin of the linear electrogyration (i.e. optical rotation induced by an applied electric field) response that was used recently to visualize the ferroaxial domains in NiTiO$_3$ ~\cite{Hayashida_2020_elecGyr}.\\
%
%
\section{Berry curvature and related responses}

\begin{figure}[b]
\begin{center}
\includegraphics[trim = 0mm 93mm 0mm 0mm, clip, width=1\linewidth]{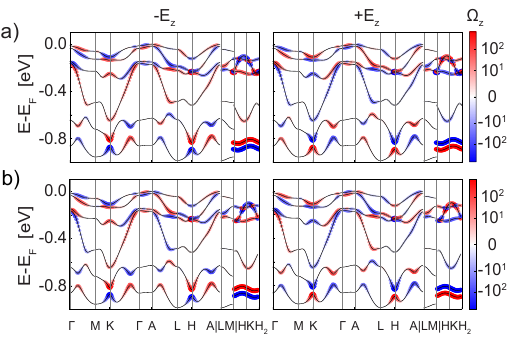}
\caption[]{Wannier-based band structure and superimposed BC component $\Omega_z$ for a) Domain 1 and b) Domain 2 with an applied electric field $E_z=\pm0.1$V/\AA. The BC is reversed for opposite fields and only partially reversed for opposite domains.}
\label{fig_bands_wannier}
\end{center}
\end{figure}

\begin{figure}[t]
\begin{center}
\includegraphics[trim = 0mm 80mm 0mm 0mm, clip, width=1\linewidth]{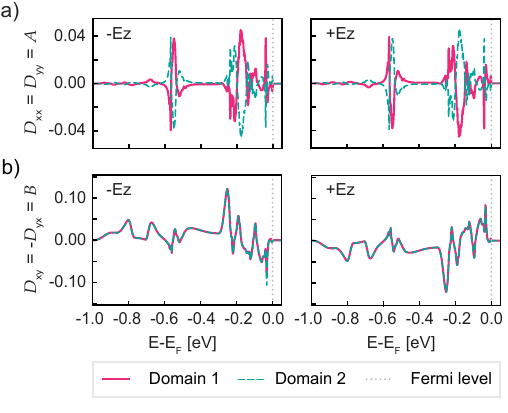}
\caption[]{Electric field induced BCD. a) For the same electric field direction, the diagonal components switch sign between ferroaxial domains. b) Off-diagonal BCD components do not switch between ferroaxial domains. Reversing the electric field for one domain yields the BCD of the opposite domain.}
\label{fig_BCD}
\end{center}
\end{figure}

The presence of TRS and IS in a crystal enforces that the BC vanishes at every $k$ point in the BZ. 
Due to this constraint, which applies to all pure ferroaxial materials, KZPO does not show BC in its ferroaxial, antiferro-chiral state. In contrast, the constraint is lifted in the induced ferri-chiral state where IS is broken by an electric field.
To characterize such a ferri-chiral state in momentum space, we next turn our attention to the BC and BCD with structural chirality imposed by ionic displacements due to a static electric field $E_z=\pm 0.1$ V/\AA.\\
First, we derived a Wannier-based tight-binding model including SOC for both ferroaxial domains, considering the isolated set of the 24 highest valence states which exhibit predominantly oxygen $p$ orbital character (see the SM). Fig.~\ref{fig_bands_wannier} shows the band structure with the computed BC component $\Omega_z$ superimposed for both ferroaxial domains and electric field orientations. While the BC is reversed everywhere upon reversal of the electric field, it only flips along some directions between opposite domains. We discuss this behavior below in the context of the BCD.
The $\Omega_z$ component is larger than $\Omega_x$ and $\Omega_y$ (reported in the SM) and the largest values occur along the H-K path ($\Omega_{x,y}$ are both zero along $H-K$ and $\Gamma-A$). In particular, these bands with the largest BC also show the largest hidden spin polarization between IS-breaking subunits~\cite{Zhang_hidden_spin:2014} without an applied electric field. \\
Next, we analyze the behavior of the BC by discussing the BCD plotted in Fig.~\ref{fig_BCD}.
The BCD tensor $D$ transforms as $D=\text{det}(S)SDS^T$~\cite{sodemann2015quantum} under the point symmetry with orthogonal matrix representation $S$.
For the $C_{3z}$ symmetry present in KZPO with an electric field along $z$, the transformation together with the zero trace condition yields
\begin{align}
    D = 
    \begin{pmatrix}
        A & B & 0\\
        -B & A & 0\\
        0 & 0 & -2A
    \end{pmatrix}.
\end{align}
We note that we obtain this form in our calculation even without imposing this symmetry. In the SM, we list the form of the BCD tensor for all eleven chiral point groups. \\
Our calculations show that within the same domain, reversing the electric field flips all BCD components.
For a fixed electric field direction, diagonal components of the BCD matrix plotted in Fig.~\ref{fig_BCD}a) switch sign between opposite domains. In contrast, the finite off-diagonal elements do not reverse between domains, as shown in Fig.~\ref{fig_BCD}b). 
\\
Our findings can be understood considering again that we induce chirality in the ferroaxial domain structures by applying an electric field parallel to the ferroaxial axis. Therefore, the domains with opposite chirality are still related by the mirror operation with normal vector along (110). When this mirror operation is applied to the BCD matrix, the off-diagonal elements are unaffected. On the other hand, a ferroaxial domain that is distorted by an electric field $+E_z$ is related by inversion to the same domain distorted by $-E_z$. Thus, BC and BCD flip sign everywhere for the same domain under opposite applied fields. This is consistent with the findings of~\cite{Joseph_2024}, in which BC and BCD were shown to flip sign between opposite enantiomers of chiral materials in enantiomorphic space groups that are related by spatial inversion.
We note that the BC and BCD are similar if instead of a 24 orbital model, we follow Ref.~\cite{Xie_KZPO} and employ a 53 orbital model including $s$ and $p$ orbitals on oxygen, phosphorus and potassium, as well as $d$ orbitals on zirconium (see the SM).\\
Before closing this section, we discuss some scenarios in which the BC effects identified in this work will manifest experimentally. First, we note that, in the presence of BCD, an electric field, $\mathcal{E}(t)=\mathrm{Re}[\mathcal{E}_{0}e^{i\omega t}]$, with frequency $\omega$, can induce a second-order oscillating electric current density, $J_{a}^{2\omega}=\chi_{abc}\mathcal{E}_{b}\mathcal{E}_{c}$. When TRS is preserved, the second-order conductivity tensor can be obtained as~\cite{sodemann2015quantum}
\begin{equation}
\chi_{abc}=\frac{\epsilon_{acd}e^3\tau}{2\hbar^2(1+i\omega\tau)}D_{bd},
\end{equation}
where $e$ is the electronic charge, $\tau$ is the relaxation time, and $\epsilon_{abc}$ is the Levi-Civita symbol. 
While KZPO is a wide-gap insulator, we envisage introduction of charge carriers either by chemical doping (for example by substituting K$^{1+}$ with Mg$^{2+}$ or Ba$^{2+}$, or Zr$^{4+}$ by  Y$^{3+}$ or Sc$^{3+}$) or ionic gating ~\cite{Ueno_gating:2008, Ueno_electrostatic_doping:2011}, and have confirmed computationally that the ferroaxial order is robust to charge doping.
(To our knowledge, very little work has been done on metallic ferroaxial materials, except for one study based on a model Hamiltonian~\cite{Hayami_metallic_ferroaxial:2023} and a recent study of TiAu$_4$ ~\cite{Jo:2026}.) 
We then evaluate the expected non-linear Hall current, taking the limit $\omega\tau \ll 1$, which is valid for typical relaxation times of the order of picoseconds, and alternating current frequency ranges from 10-1000 Hz. 
With this approximation, we find \mbox{$J_z^{2\omega}\sim \frac{e^3\tau}{2\hbar^2} D_{xy}(\mathcal{E}_x^2 + \mathcal{E}_y^2) $}. Then, taking sample dimensions of $L_x=L_y=L_z=10$ $\mu$m, an applied voltage $V_x=V_y=1$ V, an isotropic resistivity of the sample $\rho=50$ $\mu\Omega$cm, and the value of the relevant BCD component $D_{xy}\sim 0.05$ in the vicinity of the Fermi level from our first-principles calculations, we obtain the second-order Hall voltage,  $V_{z}^{2\omega} \approx 0.9$ mV along the $z$ direction. This value is well within the reach of experimental setups, and, conveniently for isolating it from other possible contributions, switches sign when the applied electric field direction is reversed.

\section{Conclusion}
In this work, we used the prototypical ferroaxial material KZPO to investigate the relation between ferroaxiality and chirality in both real and reciprocal space. Using atomic multipole decomposition, we found that the ferroaxial state in KZPO is antiferro-chiral in real space, with equal and opposite ET monopole moments associated with its previously identified IS-breaking subunits \cite{Bhowal_2024_ETD}. 
We then showed that applying an electric field uncovers this hidden chirality, inducing both a net ET monopole in real space and a net BCD in reciprocal space. Applying the field along the ferroaxial axis, so that it breaks inversion but preserves the mirror relation between ferroaxial domains, allows for particularly convenient tuning of the chiral properties, with symmetry-determined sign changes of the real and reciprocal space responses in opposite domains and for opposite field orientations. Finally, we proposed simple transport measurements, via the non-linear Hall responses, for confirming the predicted behavior. We hope that our results stimulate further exploration of the interplay between chirality and ferroaxiality, in particular experimental efforts to verify and exploit the field-induced behaviors described here.

\section{Acknowledgments} 
We thank Sayantika Bhowal and Ian Fisher for helpful discussions. This work was supported by the Swiss National Science Foundation (SNSF) under Grant Nos. 10002603 and 225790.
AN acknowledges the SNSF for sabbatical support and the Department of Science and Technology, Core Research Grant (Non-linear Hall Effect: Mechanisms and Materials, CRG/2023/000114) for funding.
Computational resources were provided by the ETH Zurich Euler cluster.

\section{Data availability}
The relevant input files of our ab initio calculations and the data supporting the findings of this work will be made publicly available upon publication of this work.  

\nocite{vaspkit:2021}

\bibliography{references}

\end{document}


\title{Supplemental Material: Hidden chiral signatures in ferroaxial K$_2$Zr(PO$_4$)$_2$}

\author{Nora Taufertsh{\"o}fer}
\affiliation{Materials Theory, ETH Z{\"u}rich, 8093 Zurich, Switzerland}
\author{Awadhesh Narayan}
\affiliation{Solid State and Structural Chemistry Unit, Indian Institute of Science, Bangalore 560012, India}
\affiliation{Materials Theory, ETH Z{\"u}rich, 8093 Zurich, Switzerland}
\author{Nicola A. Spaldin}
\affiliation{Materials Theory, ETH Z{\"u}rich, 8093 Zurich, Switzerland}

\date{\today}

\maketitle

\setcounter{section}{0}
\setcounter{equation}{0}
\setcounter{figure}{0}
\setcounter{table}{0}
\setcounter{page}{1}
\makeatletter

\renewcommand{\theequation}{S\arabic{equation}}
\renewcommand{\thefigure}{S\arabic{figure}}
\renewcommand{\thetable}{S-\Roman{table}}
\renewcommand{\bibnumfmt}[1]{[S#1]}
\renewcommand{\citenumfont}[1]{S#1}

\section{Details on the DFT calculations}

We perform ground-state density functional theory (DFT) calculations within the local density approximation (LDA) using the PAW potentials \mbox{K$_{\text{sv}}$ ([Ne]$3s^23p^64s^1$)}, \mbox{Zr$_{\text{sv}}$ ([Ar]$4s^24p^64d^35s^1$)}, \mbox{P ([Ne]$3s^23p^3$)}, and \mbox{O ([He]$2s^22p^4$)} provided with the VASP package~\cite{Kresse:1996,Kresse:1999}. We choose a $10\times10\times6$ $k-$point grid and an energy cutoff of 550 eV for the plane wave basis to converge the total energy well within 1meV per atom.
We use \textsc{vaspkit}~\cite{vaspkit:2021} to extract band structure data from the VASP output.\\
Following~\cite{Iniguez:2008}, the atomic displacement $u_{\alpha,i}$ of atom $\alpha$ in direction $i$ due to a small applied electric field $\mathbf{E}$ is determined  by
\begin{align}
    u_{\alpha,i} = \sum_n A^n d^n_{\alpha,i} \,.
\end{align}
Here, $d^n_{\alpha,i}$ is the eigenvector of the force constant matrix for mode $n$. The mode amplitude $A^n$ is given by 
\begin{align}
    \frac{1}{C^n} \sum_j p_j^n E_j 
\end{align}
with $C^n$ the eigenvalue of the force constant matrix for mode $n$ with mode polarity
\begin{align}
    p_i^n = \sum_{\alpha} \sum_j Z^{\ast}_{\alpha, ij} d_{\alpha,j}^n
\end{align}
where $Z^{\ast}$ is the Born effective charge tensor. 
In this work, we employ density functional perturbation theory in VASP to compute the Born effective charges per atom and the force constant matrix. 

\section{Details on the Wannier model and derived properties}
\begin{figure}[b!]
\begin{center}
\includegraphics[trim = 0mm 115mm 0mm 0mm, clip, width=1\linewidth]{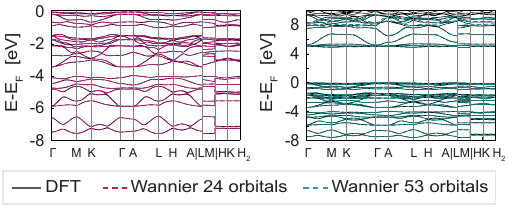}
\caption[]{DFT and Wannier-based band structures including SOC for domain 1 of KZPO with ionic distortions due to an electric field of $0.1$V/Å applied along $z$. The 24-orbital model refers to three $p$ orbitals on each oxygen site for the initial projection. The 53-orbital model refers to $s$ and $p$ orbitals on K, O, and P sites, as well as $d$ orbitals on Zr. }
\label{fig_bands_wannier_DFT}
\end{center}
\end{figure}
We employ a $6 \times 6 \times3$ $k$-point grid to build a Wannier-based tight-binding model using maximally localized Wannier functions in \textsc{Wannier90}~\cite{wannier90_Pizzi:2020}, As the valence states close to the band gap are dominated by oxygen $p$ states, we first use these orbitals to construct the wave functions. 
\begin{figure*}[!tb]
\begin{center}
\includegraphics[trim = 0mm 80mm 0mm 0mm, clip, width=1\linewidth]{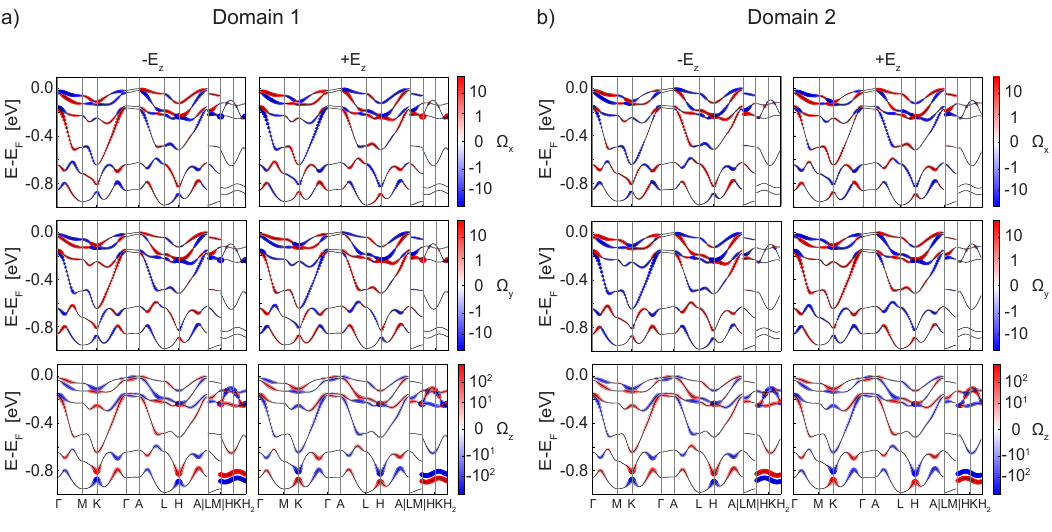}
\caption[]{BC components superimposed on the bands for the 24-orbital model for a) domain 1 and b) domain 2. }
\label{fig_BC_d1_d2_x_y_z}
\end{center}
\end{figure*}
The resulting band structure is shown in Fig.~\ref{fig_bands_wannier_DFT} on the left side for domain 1 displaced by an electric field \mbox{$E_z=0.1$ V/Å} and it matches very well with the DFT calculation.
To test the stability of the Berry curvature (BC) and Berry curvature dipole (BCD) with respect to the constructed tight-binding model, we also build a Wannier model using 53 orbitals in total, similar to~\cite{Xie_KZPO}. There, we use $s$ and $p$ orbitals on K, O, and P sites, as well as $d$ orbitals on the Zr site for the initial projection.  
The resulting band structure is shown in Fig.~\ref{fig_bands_wannier_DFT} on the right side for a limited energy range. Apart from minor deviations in the highest included conduction bands, this model reproduces well the DFT band structure.\\
We use \textsc{WannierBerri}~\cite{Tsirkin_WBerri:2021} to compute the BC and BCD for both Wannier models. Fig.~\ref{fig_BC_d1_d2_x_y_z} shows the BC based on the 24-orbital model for both domains displaced by an electric field $\pm E_z=\pm0.1$ V/Å.
\begin{figure}[h!]
\begin{center}
\includegraphics[trim = 0mm 110mm 0mm 0mm, clip, width=1\linewidth]{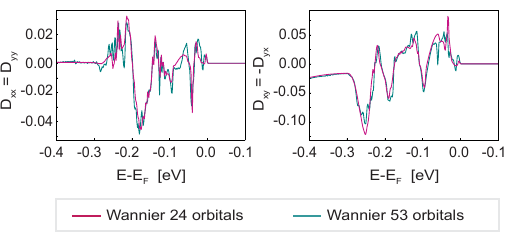}
\caption[]{BCD components based on the two different Wannier models for KZPO.}
\label{fig_BCD_comparison_ox_p_53_orb}
\end{center}
\end{figure}
To compute the BCD, we choose a $300 \times 300 \times300$ $k$-grid and \mbox{NKFFT $= 10\times10\times6$} for the Fourier transformation of the real-space Wannier Hamiltonian, which results in a reasonably well-converged BCD with the changes in the peaks $<0.005$. We report the data without applying any temperature-dependent smoothing.
Fig.~\ref{fig_BCD_comparison_ox_p_53_orb} shows the diagonal element $D_{xx}$ and the off-diagonal element $D_{xy}$ representative of the point symmetry $C_3$ in KZPO with induced ferrichirality for the 24- and 53-orbital Wannier models with the same computational settings. Although the 53-orbital model is more noisy, the BCD components match between the two models, justifying our use of the simpler 24-orbital model in the main text.
For convenience, we list the general form of the BCD tensor for all 11 chiral point groups in Table~\ref{tab:BCD}.\\
Considering the non-zero BCD components for point symmetry $C_3$ and the approximation $\omega \tau \ll1$ we find the second-order oscillating electric current densities 
\begin{align}
    J_x^{2\omega} &\sim \frac{e^3\tau}{2\hbar^2}\bigl(D_{yx}\mathcal{E}_x\mathcal{E}_z -3D_{xx} \mathcal{E}_y\mathcal{E}_z \bigr ), \nonumber \\
    J_y^{2\omega} &\sim \frac{e^3\tau}{2\hbar^2} \bigl (3D_{xx}\mathcal{E}_x\mathcal{E}_z + D_{yx} \mathcal{E}_y\mathcal{E}_z \bigr ), \\
    J_z^{2\omega} &\sim \frac{e^3\tau}{2\hbar^2} D_{xy}(\mathcal{E}_x^2+\mathcal{E}_y^2 ). \nonumber 
\end{align}
While $J_z^{2\omega}\propto D_{xy}$ switches sign when reversing the applied electric field, linear combinations of the other two components can be used to distinguish the response in experiment from other contributions. For example, $J_{x}^{2\omega}+J_{y}^{2\omega}$ switches sign between opposite applied field directions and domain orientations if $\mathcal{E}_y=-\mathcal{E}_x$. If the sample dimensions are equal, $L_x=L_y$, this switching behavior also applies to the combination $V_{x}^{2\omega}+V_{y}^{2\omega}$ of the second-order Hall voltage components.
\begin{table*}[p]
\caption{BCD tensor $D_{ab}$ for all 11 chiral point groups, derived from the symmetry constraints of each point group together with the tracelessness condition. Repeated diagonal entries are labeled $D_{\perp}$ and the off-diagonal antisymmetric pair is labeled $\pm A$. Under spatial inversion $\mathcal{P}$ (which relates enantiomorphic crystal pairs), all components of $D_{ab}$ reverse sign. The cubic groups $T$ and $O$ have $D_{ab}=0$ identically, despite being chiral and non-centrosymmetric.}
\label{tab:BCD}
\begin{ruledtabular}
\begin{tabular}{lllc}
\textbf{Point group} &
\textbf{H--M} &
\textbf{Crystal system} &
\textbf{Tensor $D_{ab}$} \\
\midrule

$C_1$ & $1$ & Triclinic &
$\begin{pmatrix}
D_{xx}  & D_{xy} & D_{xz} \\
D_{yx}  & D_{yy} & D_{yz} \\
D_{zx}  & D_{zy} & -(D_{xx}+D_{yy})
\end{pmatrix}$ \\[16pt]

\midrule

$C_2$ & $2$ & Monoclinic &
$\begin{pmatrix}
D_{xx} & D_{xy} & 0 \\
D_{yx} & D_{yy} & 0 \\
0      & 0      & -(D_{xx}+D_{yy})
\end{pmatrix}$ \\[16pt]

\midrule

$D_2$ & $222$ & Orthorhombic &
$\begin{pmatrix}
D_{xx} & 0      & 0 \\
0      & D_{yy} & 0 \\
0      & 0      & -(D_{xx}+D_{yy})
\end{pmatrix}$ \\[16pt]

\midrule

$C_3$ & $3$ & Trigonal &
$\begin{pmatrix}
D_{\perp} &  A  & 0 \\
-A        & D_{\perp} & 0 \\
0         &  0  & -2D_{\perp}
\end{pmatrix}$ \\[16pt]

$D_3$ & $32$ & Trigonal &
$\begin{pmatrix}
D_{\perp} & 0 & 0 \\
0 & D_{\perp} & 0 \\
0 & 0 & -2D_{\perp}
\end{pmatrix}$ \\[16pt]

\midrule

$C_4$ & $4$ & Tetragonal &
$\begin{pmatrix}
D_{\perp} &  A  & 0 \\
-A        & D_{\perp} & 0 \\
0         &  0  & -2D_{\perp}
\end{pmatrix}$ \\[16pt]

$D_4$ & $422$ & Tetragonal &
$\begin{pmatrix}
D_{\perp} & 0 & 0 \\
0 & D_{\perp} & 0 \\
0 & 0 & -2D_{\perp}
\end{pmatrix}$ \\[16pt]

\midrule

$C_6$ & $6$ & Hexagonal &
$\begin{pmatrix}
D_{\perp} &  A  & 0 \\
-A        & D_{\perp} & 0 \\
0         &  0  & -2D_{\perp}
\end{pmatrix}$ \\[16pt]

$D_6$ & $622$ & Hexagonal &
$\begin{pmatrix}
D_{\perp} & 0 & 0 \\
0 & D_{\perp} & 0 \\
0 & 0 & -2D_{\perp}
\end{pmatrix}$ \\[16pt]

\midrule

$T$ & $23$ & Cubic &
$\begin{pmatrix}
0 & 0 & 0 \\
0 & 0 & 0 \\
0 & 0 & 0
\end{pmatrix}$ \\[16pt]

$O$ & $432$ & Cubic &
$\begin{pmatrix}
0 & 0 & 0 \\
0 & 0 & 0 \\
0 & 0 & 0
\end{pmatrix}$ \\[8pt]

\bottomrule
\end{tabular}
\end{ruledtabular}
\end{table*}

\section{Spin polarization of inversion-breaking subunits}

\begin{figure}[t]
\begin{center}
\includegraphics[trim = 0mm 92mm 0mm 0mm, clip, width=1\linewidth]{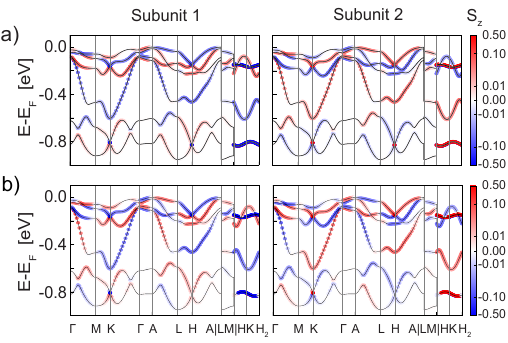}
\caption[]{Spin polarization of the two inversion-breaking subunits for a) domain 1 and b) domain 2 without an electric field applied.}
\label{fig_spin_pol}
\end{center}
\end{figure}

Inversion-breaking subunits in an overall inversion-symmetric crystal structure yield a hidden opposite spin polarization~\cite{Zhang_hidden_spin:2014}, which was previously demonstrated for KZPO~\cite{Bhowal_2024_ETD}.
We reproduced this result and found that the spin polarization between subunits is strongest along the $H-K-H_2$ path for two Kramers' pairs that are only slightly split by SOC and show a similar, flat dispersion (see Fig.~\ref{fig_spin_pol}).
Note that, as all bands are 2-fold degenerate due to Kramers' theorem, we sum the spin polarization over these degenerate bands, respectively. The almost 4-fold degenerate bands along the high-symmetry path $H-K-H_2$ are not summed over, as they are slightly split by SOC. These bands appear almost on top of each other in Fig.~\ref{fig_spin_pol} and their spin polarization is the strongest with almost $\pm1/2$. 
%
These respectively two Kramers' pairs along $H-K-H_2$ that are very close in energy and yield the largest spin polarization also show a finite BC, even if inversion symmetry is restored in the limit $\mathbf{E} \rightarrow 0$, as long as  SOC is included. In this case, the BC becomes zero at each $k$-point only if the sum over both Kramer's pairs, i.e. over four bands, is considered.
%
%

\bibliography{references}